\documentclass[conference, 10pt]{IEEEtran}

\usepackage{cite}
\usepackage[cmex10]{amsmath}
\usepackage{amssymb}
\usepackage{url}
\usepackage{graphicx}
\usepackage{color}
\usepackage{placeins}
\usepackage{float}
\usepackage{tabularx,colortbl}
\usepackage{ifthen}

\def\Rob#1{\textcolor{black}{#1}} %Roberto blue  

\makeatletter

\newcounter{author}
\renewcommand{\author}[2][]{
   \stepcounter{author}
   \@namedef{author@\theauthor}{#2}
   \@namedef{authorlabel@\theauthor}{#1}
}

\newcounter{address}
\newcommand{\address}[2][]{
   \stepcounter{address}
   \@namedef{address@\theaddress}{#2}
   \@namedef{addresslabel@\theaddress}{#1}
}

\newcommand{\alsep}{and}

\def\newmaketitle{\par%
  \begingroup%
  \normalfont%
  \def\thefootnote{}%  the \thanks{} mark type is empty
  \def\footnotemark{}% and kill space from \thanks within author
  \let\@makefnmark\relax% V1.7, must *really* kill footnotemark to remove all \textsuperscript spacing as well.
  \footnotesize%       equal spacing between thanks lines
  \footnotesep 0.7\baselineskip%see global setting of \footnotesep for more info
  \normalsize%
  \twocolumn[\thenewmaketitle\@IEEEaftertitletext]%
  \if@IEEEusingpubid
     \enlargethispage{-\@IEEEpubidpullup}%
  \fi
  \endgroup
  \setcounter{footnote}{0}\let\maketitle\relax\let\@maketitle\relax
  \gdef\@thanks{}%
  \let\thanks\relax}

\def\thenewmaketitle{
  \newpage
  \begin{center}%
    \vskip0.2em{\Huge\@IEEEcompsoconly{\sffamily}\@IEEEcompsocconfonly{\normalfont\normalsize\vskip 2\@IEEEnormalsizeunitybaselineskip
   \bfseries\large}\@title\par}\vskip1.0em\par%
    \vspace{1ex}
    \newcounter{c@author}
    \newcounter{c@tmp}
    \ifthenelse{\value{author}=2}{%
      \newcommand{\liand}{ and }}{%
      \newcommand{\liand}{, and }}
    \ifthenelse{\value{address}<2}{%
      \@nameuse{author@1}%
      \stepcounter{c@author}%
      \whiledo{\value{c@author}<\value{author}}{%
        \setcounter{c@tmp}{\value{author}}%
        \addtocounter{c@tmp}{-\value{c@author}}%
        \ifthenelse{\value{c@tmp}=1}{%
          \renewcommand{\alsep}{\liand}}{\renewcommand{\alsep}{, }}%
        \stepcounter{c@author}\alsep \@nameuse{author@\thec@author}}\\%
    }
    {%Add address references after the author's name
      \@nameuse{author@1}${}^{(\ref{\@nameuse{authorlabel@1}})}$%
      \stepcounter{c@author}%
      \whiledo{\value{c@author}<\value{author}}{%
      \setcounter{c@tmp}{\value{author}}%
      \addtocounter{c@tmp}{-\value{c@author}}%
      \ifthenelse{\value{c@tmp}=1}{%
        \renewcommand{\alsep}{\liand}}{\renewcommand{\alsep}{, }}%
      \stepcounter{c@author}\alsep \@nameuse{author@\thec@author}%
        ${}^{(\ref{\@nameuse{authorlabel@\thec@author}})}$%
      }
    }
    \vspace{0.2ex}

    \ifthenelse{\value{address}>0}{%
      \ifthenelse{\value{address}=1}{
        {\@nameuse{address@1}}
      }
      {%Output the addresses as an enumerated list
        \newcounter{c@address}

        \begin{center}
        \whiledo{\value{c@address}<\value{address}}
        {
          \refstepcounter{c@address}
            ${}^{(\thec@address)}$\,%
              \label{\@nameuse{addresslabel@\thec@address}}%
              \@nameuse{address@\thec@address}\\ %
        }
        \end{center}
      } % end of the address creation ifthenelse block
    }
    {
      \relax
    }
  \end{center}
}

\makeatother

\title{Design and Optimization of Reconfigurable Intelligent Surfaces Using the PEEC Method}

\author[org1]{Giuseppe Pettanice}
\author[org2]{Marco Di Renzo}
\author[org1]{Roberto Valentini}
\author[org2]{Sumin Jeong}
\author[org1]{Piergiuseppe Di Marco}
\author[org1]{Fortunato Santucci}
\author[org1]{Daniele Romano}
\author[org1]{Giulio Antonini}

\address[org1]{University of L’Aquila, 67100 L’Aquila, Italy}
\address[org2]{Universit\'e Paris-Saclay, CNRS, CentraleSup\'elec, Laboratoire des Signaux et Syst\`emes, France.}

\begin{document}

\newmaketitle

\begin{abstract}
The design and optimization of Reconfigurable Intelligent Surfaces (RISs) are key challenges for future wireless communication systems. RISs are devices that can manipulate electromagnetic (EM) waves in a programmable way, thus enhancing the performance and efficiency of wireless links. To achieve this goal, it is essential to have reliable EM models that can capture the behavior of RISs in different scenarios. This work demonstrates that the Partial Elements Equivalent Circuit (PEEC) method is a powerful tool for EM analysis of RIS-aided wireless links. It might also be integrated with optimization algorithms in order to optimize wireless communication networks.
\end{abstract}

\section{Introduction}\label{sec:intro}
Reconfigurable intelligent surfaces (RISs) are a novel technology that can enhance wireless communication systems by \Rob{smartly} manipulating the electromagnetic waves in the environment. An RIS is composed of a large number of nearly-passive scattering elements, that can be programmed to adjust their reflection, absorption, or refraction properties. By doing so, an RIS can create favorable propagation conditions for wireless signals, such as beamforming, interference cancellation, or secure communications \cite{DiRe20etal}.
%One of the main advantages of RIS is that it can achieve signal focusing of the scattered field, which means that it can concentrate the wireless energy in a desired direction or location. 
One of the main advantages of RISs \Rob{relies in the capability of} focusing the scattered field, \Rob{meaning} that it can concentrate the wireless energy in a desired direction or location. 
%This can improve the performance of wireless communication systems in terms of coverage, capacity, energy efficiency, and security. However, to achieve this goal, communication models that take into account the physics and electromagnetic (EM) features of the RIS elements that scatter the waves are needed to study and improve wireless systems with RIS. 
This is an enabler for enhancing the performance of wireless networks in terms of coverage, capacity, energy efficiency, and security. However, to achieve this goal, \Rob{accurate} communication models that take into account the physics and electromagnetic (EM) features of the RIS elements are \Rob{of paramount importance for optimizing RIS-aided systems.} \Rob{Importantly, this calls for EM-consistent models that accurately describe the wave propagation and RIS elements, at a tractable complexity \cite{Grad21DR,Petta23},\cite{MDR24}.}
%These models should be close to reality, precise, and manageable. 
%This means we need EM models that can explain well how the waves propagate and how the RIS elements interact \cite{Grad21DR,Petta23}.

%Moreover, the RIS must be optimized according to the wireless links' channel state information (CSI). 
Moreover, \Rob{a crucial aspect is to optimally configure} the RIS according to the channel state information (CSI).
The optimization problem consists of finding the optimal configuration of the RIS elements that maximizes a certain objective function, such as the received signal power, the signal-to-noise ratio, or the secrecy rate.
The optimization of RISs is a challenging task for several reasons. First, the RIS elements have discrete and finite states, which makes the optimization problem non-convex and combinatorial. Second, the CSI of RIS-empowered wireless links is difficult to obtain since an RIS do not have any active components for \Rob{channel probing.} %or feedback. 
Third, the optimization problem may involve multiple RISs, multiple users, and multiple objectives, which increases the complexity and the dimensionality of the problem. An \Rob{EM-consistent} optimization algorithm was introduced in \cite{HAS23}.

\Rob{In this work, we propose the Partial Elements Equivalent Circuit (PEEC) method \cite{Ru17AJ} as an effective and powerful tool for the EM characterization of RIS-aided communication channels. Moreover, we stress the importance of optimizing the RIS configuration by considering the RIS radiation pattern \cite{MDR22}. Finally, to demonstrate the effectiveness of the PEEC method, we compare the obtained results with those obtained with the commercial software Feko.}

%This work aims to prove the versatility of the Partial Elements Equivalent Circuit (PEEC) method \Rob{[citation here?]} in the EM modeling of the communication channel. We also emphasize the importance of optimizing the RIS by showing the radiation pattern of an optimized RIS and comparing the results of the PEEC method with those of the commercial software Feko.

\section{The PEEC Method}\label{sec:sysMOD}
The PEEC method is a powerful computational technique for analyzing EM phenomena in various systems and structures \cite{Ru17AJ}. 
It solves the Electric Field Integral Equation (EFIE) and the Continuity Equation (CE), but, differently from the Method of Moments (MoM) \cite{Harr93}, the two equations are kept
separate and, thus, the currents and charges (or potentials) are kept separate as well. The geometry is meshed using volumes and surfaces, which can be hexahedra/quadrilaterals or tetrahedra/triangles. The electric currents are assumed to flow through the elementary volumes, while the electric charges are assumed to exist on the elementary surfaces of the mesh. This allows for identifying an equivalent circuit to which the Kirchhoff voltage and current laws are enforced. The interactions of the electric field are described by the coefficients of the potential $\mathbf{P}$, and the interactions of the magnetic field are described by the partial inductances $\mathbf{L}_P$. The conductor losses are captured by resistances and the polarization currents in dielectrics by capacitances \cite{Ru17AJ}. The main advantage of a circuit representation for an electromagnetic problem is the possibility of adding linear and nonlinear components, and the possibility of solving the resulting circuit in both the frequency and time domains. The resulting PEEC equivalent circuit can then be analyzed using various techniques such as the Modified Nodal Analysis (MNA) \cite{Ru17AJ} in the frequency domain, as follows:
\begin{equation}\label{eq:FD_PEEC}
\left[
\begin{array}{cc}
    \boldsymbol{\mathcal Z}(s) + s \mathbf L_p & - \mathbf {A^T} \\
    \mathbf A & s \mathbf P^{-1} + \mathbf{Y}_{\ell e}(s) 
\end{array} 
\right] 
\left[ \begin{array}{c}
       \mathbf I (s)\\ 
       \boldsymbol{\Phi} (s)    
\end{array} \right] = 
\left[ \begin{array}{c}
       \mathbf {V_s}(s)\\
       \mathbf {I_s}(s)
\end{array} \right]
\end{equation}
where $\boldsymbol{\mathcal Z}(s)$, represents the impedance of the elementary volumes of conductors or dielectrics. The lumped admittance matrix, $\mathbf Y_{\ell e}$, encompasses all the lumped admittances linked to the nodes of the equivalent circuit, and $\mathbf A$ is the incidence matrix \cite[Eq. (13.6)]{Ru17AJ}.
The unknown vectors of node potentials and branch electric currents, denoted by $ \boldsymbol{\Phi}(s)$ and $\mathbf I (s)$ respectively, can be determined from the voltage and current sources, $\mathbf V_s(s)$ and $\mathbf I_s(s)$ respectively.
%With the understanding of $\boldsymbol{\Phi}(s)$ and $\mathbf I (s)$, it becomes feasible to calculate the port voltages and currents.
From the model, it is possible to calculate the impedance matrix of the entire system, $\mathbf Z_{sys}$, which accounts for the interactions between the transmitters, receivers, scatterers, and RIS. 
The resulting analysis provides insights on the system behavior, including the channel gain, impedance matching, and signal integrity.

\section{RIS Optimization}
\label{sec:opt}
The goal of RIS optimization is to maximize the achievable rate, which represents the highest rate at which data can be reliably transmitted over a wireless link. 
The RIS terminations affect the achievable rate, which can be written as a function of them. 
We adopt the algorithm introduced in \cite{HAS23}, which computes the optimal load impedances one by one in an iterative manner. The algorithm uses three key mathematical techniques, such as Sherman-Morrison’s formula, Sylvester’s theorem, and Gram-Schmidt’s method, to simplify the optimization problem and find a closed-form solution for the optimal impedances at each iteration of the algorithm. This makes the algorithm faster and more efficient than existing methods, as it needs fewer iterations and less time to converge.

%
% \vspace{-0.1cm}
%
\section{Numerical Results}\label{sec:numres}
\begin{figure}[!t]
\centerline{{
\includegraphics[width=1\columnwidth]{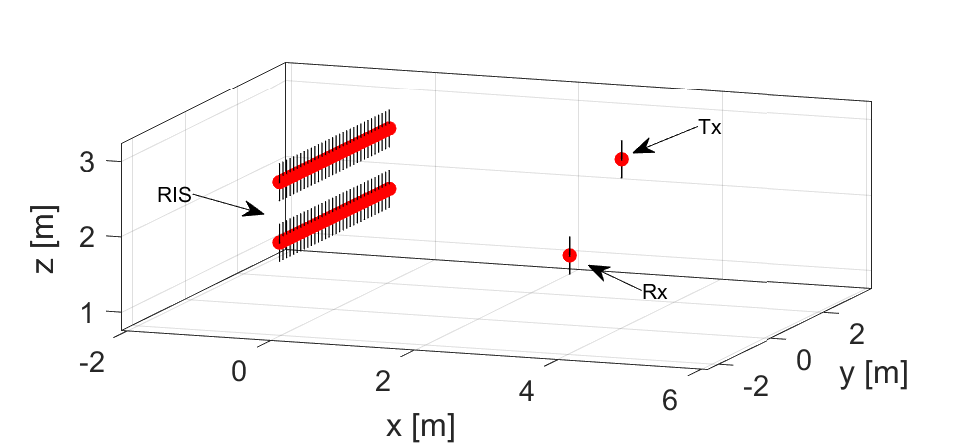}}}
\caption{System configuration. }
\label{fig:sys_config}
\end{figure}

\begin{figure}[!t]
\centerline{{
\includegraphics[width=1\columnwidth]{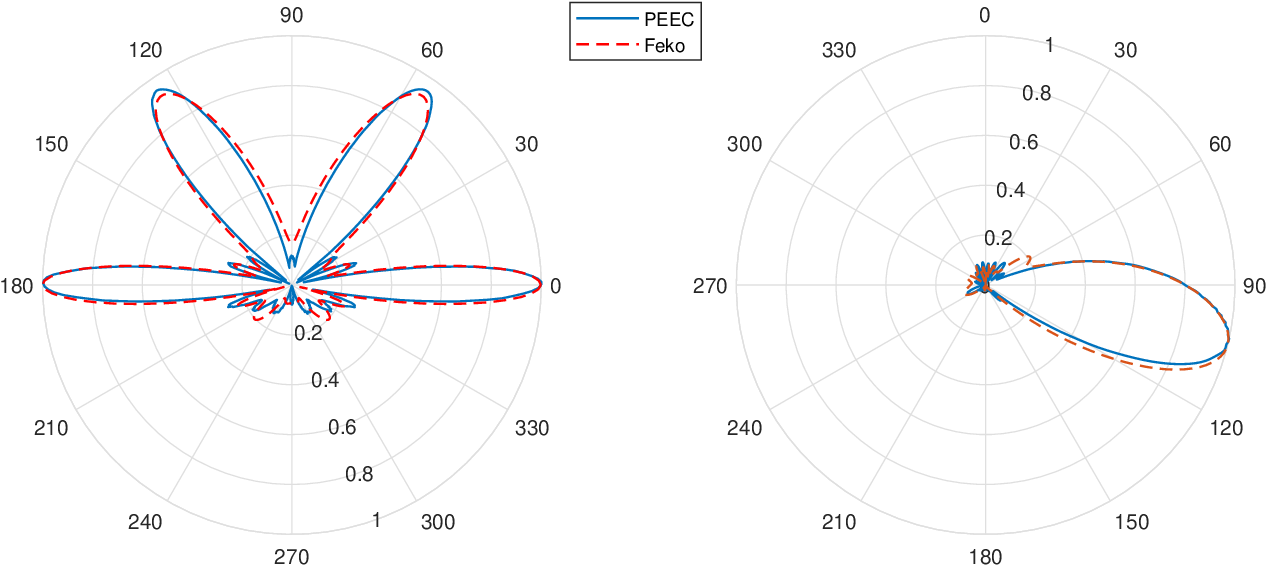}}}
\caption{Normalized radiation pattern: $\phi-$plane (left); $\theta-$plane (right). }
\label{fig:rp}
\end{figure}

We consider \Rob{the} scenario \Rob{illustrated in Fig. \ref{fig:sys_config}}, consisting of thin-wire dipoles. The transmitter is centered in $\mathbf r_{Tx} = [4\ 0\ 3]$ m; the receiver is centered in $\mathbf r_{Rx}=[2\ 3.46\ 1]$ m, and the RIS, consisting of an array of dipoles, is centered in $\mathbf r_{RIS}=[0\  0\ 2]$ m. The array is made up of two rows of $32$ dipoles arranged on the $yz$-plane, and the elements are spaced by $d_y=\lambda/8$ and $d_z=3\lambda/4$, with $\lambda$ being the wavelength. The dipoles are parallel to the $z-$axis, and their length is $\lambda/2$. The resonance frequency is $28$ GHz, corresponding to $\lambda \cong 1$ cm. %The described setup is illustrated in Fig. \ref{fig:sys_config}.
The red dots in Fig. \ref{fig:sys_config} indicate the ports for feeding and terminating the dipoles.
Figure \ref{fig:rp} shows how the scattered field of the RIS is optimized in both the azimuth $(\phi)$ and the elevation $(\theta)$ planes. With respect to the RIS center, the receiver is oriented towards $60^{\circ}$ in the $\phi-$plane, and approximately towards $104^{\circ}$ in the $\theta-$plane. The results show a good agreement between the PEEC model and Feko.

%
% \vspace{0.35cm}
%
\section{Conclusion}\label{sec:con}
\Rob{In this study, we proposed the PEEC method as an effective tool for EM modeling in RIS-aided wireless systems. We showed that the PEEC formulation can be efficiently integrated in existing optimization algorithms and validated the results by using a commercial EM simulator.}
%In this study, our results showed the potential of the PEEC method used for the electromagnetic modeling of communication systems. It might be well-integrated with optimization algorithms. The study is validated by comparing PEEC results with those of commercial software, e.g., Feko.
%
\vspace{-0.1cm}
\section*{ACKNOWLEDGEMENT}
Giuseppe Pettanice acknowledges the Filauro Foundation for supporting him while staying at the Universit\'e Paris-Saclay, CentraleSup\'elec, 91192 Gif-sur-Yvette, France. 
The work of M. Di Renzo was supported in part by the European Commission through the Horizon Europe project titled COVER under grant agreement number 101086228, the Horizon Europe project titled UNITE under grant agreement number 101129618, and the Horizon Europe project titled INSTINCT under grant agreement number 101139161, as well as by the Agence Nationale de la Recherche through the France 2030 project titled ANR-PEPR Networks of the Future under grant agreement NF-PERSEUS, 22-PEFT-004, and ANR-CHIST-ERA project titled PASSIONATE under grant agreement CHIST-ERA-22-WAI-04 through ANR-23-CHR4-0003-01.
This work was also partially supported by the Centre of Excellence on Connected, Geo-Localized, and Cyber Secure Vehicles (Ex-EMERGE), funded by the Italian Government under CIPE Resolution 70/2017. 
\vspace{-0.1cm}
%
% \vspace{-0.0cm}
%
% used to balance the columns on the last page adjust value as needed


%\bibliographystyle{IEEEtran}
\bibliographystyle{IEEEtran}
\bibliography{bibl/Abbr12, bibl/Ref2, bibl/Ref14}

\end{document}